\documentclass[apjl, iop]{emulateapj}

\newcommand \Hbeta {\ifmmode {\rm H}\beta \else H$\beta$\fi}
\newcommand \hb    {\ifmmode {\rm H}\beta \else H$\beta$\fi}
\newcommand  \mgii  {\ifmmode {\rm Mg}{\textsc{ii}} \else Mg\,{\sc ii}\fi}
\newcommand  \MGII  {\ifmmode {\rm Mg}\,{\sc ii}\,\lambda2798 \else Mg\,{\sc ii}\,$\lambda2798$\fi}
\newcommand  \siiv  {\ifmmode {\rm Si}\, {\sc iv}\ \else Si\,{\sc iv}\fi}
\newcommand  \SIIV  {\ifmmode {\rm Si}\,{\sc iv}\,\lambda1399 \else Si\,{\sc iv}\,$\lambda1399$\fi}
\newcommand  \civ  {\ifmmode {\rm C}\, {\sc iv}\ \else C\,{\sc iv}\fi}
\newcommand  \CIV  {\ifmmode {\rm C}\,{\sc iv}\,\lambda1549 \else C\,{\sc iv}\,$\lambda1549$\fi}
\newcommand  \NV  {\ifmmode {\rm N}\,{\sc v}\,\lambda1240 \else N\,{\sc v}\,$\lambda1240$\fi}
\newcommand  \nv  {\ifmmode {\rm N}\,{\sc v}\ \else N\,{\sc v}\fi}
\newcommand  \LyA  {\ifmmode {\rm Lyman}\,{\sc $\alpha$}\,\lambda1216 \else Lyman\,{\sc $\alpha$}\,$\lambda1216$\fi}
\newcommand  \feii     {Fe\,{\sc ii}}
\newcommand  \feiii     {Fe\,{\sc iii}}
\newcommand  \aliii  {\ifmmode {\rm Al}{\textsc{iii}} \else Al\,{\sc iii}\fi}
\newcommand  \ALIII  {\ifmmode {\rm Al}\,{\sc iii}\,\lambda1857 \else Al\,{\sc iii}\,$\lambda1857$\fi}

\newcommand  \CIII  {\ifmmode {\rm C}\,{\sc iii]}\,\lambda1909 \else C\,{\sc iii]}\,$\lambda1909$\fi}
\newcommand  \oi    {\ifmmode \left[{\rm O}\,{\textsc i}\right] \else [O\,{\sc i}]\fi}
\newcommand  \OI    {\ifmmode \left[{\rm O}\,{\textsc i}\right]\,\lambda6300 \else [O\,{\sc i}]$\,\lambda6300$ \fi}
\newcommand  \hei     {He\,{\sc i*}}
\newcommand  \oii   {\ifmmode \left[{\rm O}\,{\textsc ii}\right] \else [O\,{\sc ii}]\fi}
\newcommand  \OII   {\ifmmode \left[{\rm O}\,{\textsc ii}\right]\,\lambda3727 \else [O\,{\sc ii}]\,$\lambda3727$ \fi}
\newcommand  \oiii  {\ifmmode \left[{\rm O}\,{\textsc iii}\right] \else [O\,{\sc iii}]\fi}
\newcommand  \OIII  {\ifmmode \left[{\rm O}\,{\textsc iii}\right]\,\lambda5007 \else [O\,{\sc iii}]\,$\lambda5007$\fi}
\newcommand  \ovi    {\ifmmode \left[{\rm O}\,{\textsc vi}\right] \else O\,{\sc vi}\fi}
\newcommand  \neiii   {\ifmmode \left[{\rm Ne}\,{\textsc iii}\right] \else [Ne\,{\sc iii}]\fi}
\newcommand  \nev   {\ifmmode \left[{\rm Ne}\,{\textsc v}\right] \else [Ne\,{\sc v}]\fi}

\newcommand{\kms}{\ifmmode {\rm km\,s}^{-1} \else km\,s$^{-1}$ \fi}

\usepackage{ulem}

\shorttitle{\mgii\ BAL transitions}
\shortauthors{Yi et al.}

\begin{document}


\title{ Multi-epoch spectroscopy of \mgii\ broad absorption line transitions } 

\author{Weimin Yi\altaffilmark{1,2,3} and John Timlin\altaffilmark{2}   }


\altaffiltext{1}{Yunnan Observatories, Kunming, 650216, China}
\altaffiltext{2}{Department of Astronomy \& Astrophysics, The Pennsylvania State University, 525 Davey Lab, University Park, PA 16802, USA}  
\altaffiltext{3}{Key Laboratory for the Structure and Evolution of Celestial Objects, Chinese Academy of Sciences, Kunming 650216, China}

\begin{abstract}
Built upon a sample of 134 quasars that was dedicated to a systematic study of \mgii-BAL variability from \cite{Yi19a}, we investigate these quasars showing \mgii-BAL disappearance or emergence  with the aid of at least three epoch optical  spectra sampled more than 15 yr in the observed frame. We identified 3/3 quasars undergoing pristine/tentative BAL transformations. The incidence of pristine BAL transformations in the sample is therefore derived to be 2.2$_{-1.2}^{+2.2}$\%, consistent with that of high-ionization BAL transformations from the literature. Adopting an average \mgii-BAL disappearance timescale of rest-frame 6.89 yr among the six quasars, the average characteristic  lifetime of \mgii\ BALs in the sample is constrained to be $>$160 yr along our line of sight. There is a diversity of BAL-profile variability observed in the six quasars, probably reflecting a variety of mechanisms at work. Our investigations of \mgii-BAL transitions, combined with observational studies of BAL transitions from the literature, imply an overall  FeLoBAL/LoBAL$\rightarrow$HiBAL/non-BAL transformation sequence along with a decrease in reddening. This sequence is consistent with the evacuation models for the origin of commonly seen blue quasars, in which LoBAL quasars are in a shorted-lived, blowout phase. 
\end{abstract}


\section{Introduction} \label{introduction}

Broad absorption line (BAL) quasars, which make up $\sim$15\%--40\% among the quasar population (e.g., \citealp{Gibson09,Allen11,Paris17}), are usually classified into three subtypes, depending on the ionization potentials of the ions present in their absorption troughs. The majority are classified as high-ionization BAL quasars (HiBALs), which usually contain absorption troughs caused by ionic species such as \civ, \nv, and \ovi. Low-ionization BAL quasars (LoBALs, $\sim$10\% of the BAL-quasar population, e.g., \citealp{Trump06}) are characterized by the presence of additional low-ionization species (e.g., \mgii, \aliii) in their spectra. Fe low-ionization BAL quasars (FeLoBALs) show prominent \feii\ and/or \feiii\ in addition to other low-ionization species. 
BAL phenomena are characterized by remarkable absorption features imprinted on spectra, with both trough width and maximum trough velocity larger than 2000 \kms, unambiguously signaling quasar-driven winds. Mini-BALs, which have trough width larger than 500 \kms\ and less than 2000 \kms, are likely another manifestation of quasar-driven outflows (\citealp{Hamann04,Moravec17}). Together with intrinsic narrow absorption line (NAL) phenomena that are somewhat difficult for identification, about 70\% quasars have outflow signatures traced by one or more of the above three manifestations imprinted on spectra (\citealp{Hamann12}).

LoBAL quasars are often interpreted as a short-lived stage in which the central engine, namely supermassive black hole (SMBH), is evacuating large amounts of material and energy from the circumnuclear region in the context of an evolutionary picture (e.g., \citealp{Sanders88,Urrutia09,Zubovas13}). Mounting evidence with kpc-scale BAL outflows has been found in this population and the inferred kinetic luminosity of BAL outflows tend to be higher than 1\%, which are potentially capable of triggering quasar feedback (e.g., \citealp{Moe09,Hopkins10,Borguet13,Yi17,Arav18}).

While the vast majority of quasars cannot be spatially resolved with current technology, variability diagnostics have been widely used to explore the quasar inner regions. 
BAL variability has been explained primarily by changes in the ionization state of outflowing gas, transverse motions across our line of sight (LOS), or a combination of both mechanisms (e.g., \citealp{Cap11,Filizak13,Wang15,He17,Yi19a}). 
Previous studies based on large samples have found that BAL variability appears to be ubiquitous over different-sampling timescales (from days to years, e.g., \citealp{Filizak13,Hemler19}), with variability in amplitude becoming larger on longer sampling timescales both in HiBAL and LoBAL quasars (\citealp{Filizak13,Yi19a}). 
However, a full analysis of BAL variability is complicated, as one needs to consider partial covering and saturation effects, overlapping absorption troughs, inhomogeneous absorbers and background emitting sources, rapid BAL variability, velocity-/density-/ionization-/wavelength-dependent covering factors etc (\citealp{Hall02,Arav12,Hamann19}). 
In a technical sense, the analysis of BAL variability depends on sampling timescale and cadence, data quality (both in spectral S/N and resolution), the availability of multi-ion BALs and/or multi-wavelength data etc. 
Therefore, BAL-variability studies based on relatively low-resolution and low-S/N spectra can reveal only an overall change across a wide BAL trough via the commonly adopted equivalent width (EW). Without tracing detailed BAL-profile changes, the analysis may suffer large uncertainties, especially for individual quasars.

One of the great challenges in studying BALs is to ascertain whether all velocity components appearing in a BAL trough arise from a closely similar location. This is because an observed BAL imprinted on a spectrum signals all absorption components along our LOS. A good example, as revealed by \citet{Arav15}, the \civ\ BAL trough in NGC 5548 consists of at least six different-velocity components at largely different locations along our LOS. Another example comes from the LoBAL quasar LBQS 1206+1052, which has a BAL trough consisting of two different-velocity components separated by $\sim$700 km~s$^{-1}$ (\citealp{Ji12}). Based on high-quality spectra, the two velocity components were found to be located at $\sim$1 pc and $\sim$500 pc, respectively (e.g., \citealp{Sun2017, Miller2018}). In fact, the vast majority of SDSS BAL quasars do not have high-quality, multi-wavelength spectra and BALs often partially cover their background light sources, making the identification of different-velocity components in a BAL trough impossible. A complete BAL disappearance or emergence event, however, naturally addresses this challenge even with relatively low-quality spectra (See Section~\ref{discussion}). Such events have been reported mostly in HiBALs (\citealp{Hamann08,Filizak12,Wang15,McGraw2017,Stern2017, DeCicco18,Rogerson18,Sameer19}); by contrast, only three MgII-BAL disappearance and/or emergence cases have been reported to date (\citealp{Junkkarinen2001,Vivek12,Yi19b}). 

In this work, we conduct a systematic analysis of \mgii-BAL disappearance and emergence built upon the \mgii-BAL sample consisting of 134 quasars in \citet{Yi19a} with archived multi-epoch SDSS spectra over more than 10 yr, complemented by  additional spectroscopic observations. We identified six quasars having \mgii-BAL disappearance in the sample. Using three or more spectroscopic epochs that are well separated in time for each BAL quasar in the six quasars, we aim to gain additional insight into the inner structures and underlying physics with the aid of BAL transitions. 
Throughout this work, we clarify that both obscurer and absorber are the different parts of a disk wind; specifically, the former represents the base part of the disk wind that lies mostly between broad line region (BLR) and continuum region while the latter refers to the extended and possibly accelerated part of the disk wind.

\section{Observations and data reduction}

We obtained additional spectra for the six quasars with detection of BAL transitions using the Low-Resolution Spectrograph-2 (LRS2; \citealp{Chonis14}) mounted on the Hobby-Eberly Telescope (HET; \citealp{Ramsey98}) and the Yunnan Faint Object Spectrograph and Camera (YFOSC) mounted on the Lijiang 2.4m Telescope (LJT; \citealp{Fan15,WangC2019}). The Fast Turnaround program (program ID: 2018B-FT-214) from the Gemini Observatory allows us to obtain an additional spectrum for the faintest quasar (J082747) in our sample, which shows both BAL disappearance and emergence. We tabulate the detailed observations for the six LoBAL QSOs in Table~\ref{table1}, in which the last-epoch spectrum for each quasar was obtained from our follow-up observations. 

\begin{figure*}
\center
\resizebox{6in}{!}{\includegraphics{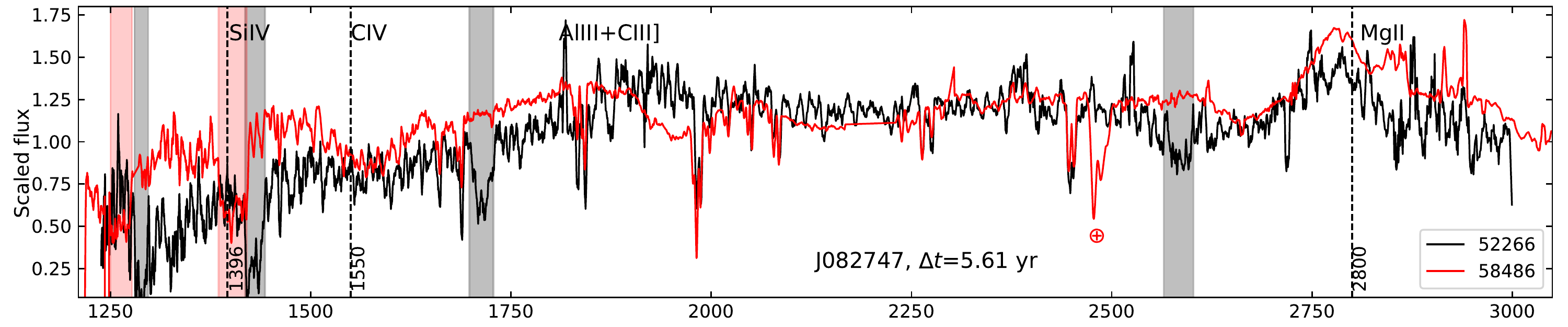}} 
\resizebox{6in}{!}{\includegraphics{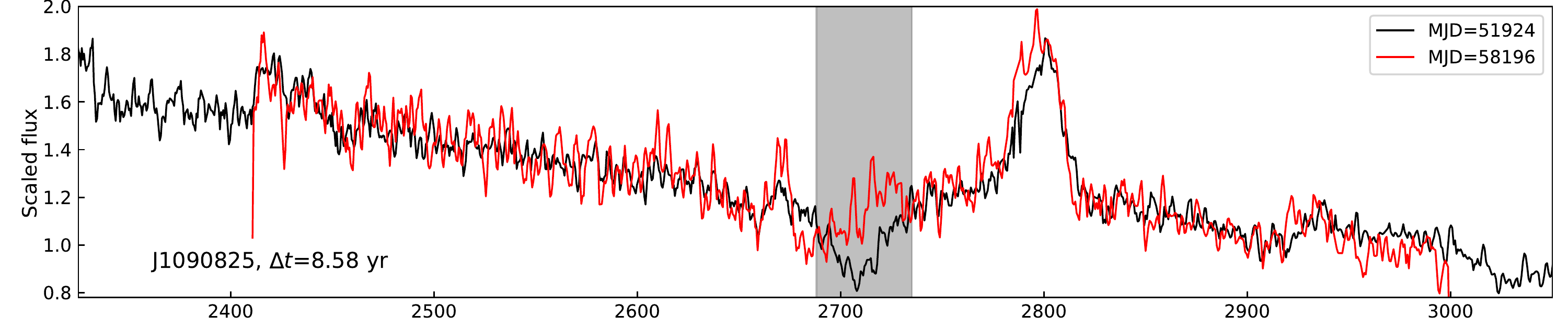}} 
\resizebox{6in}{!}{\includegraphics{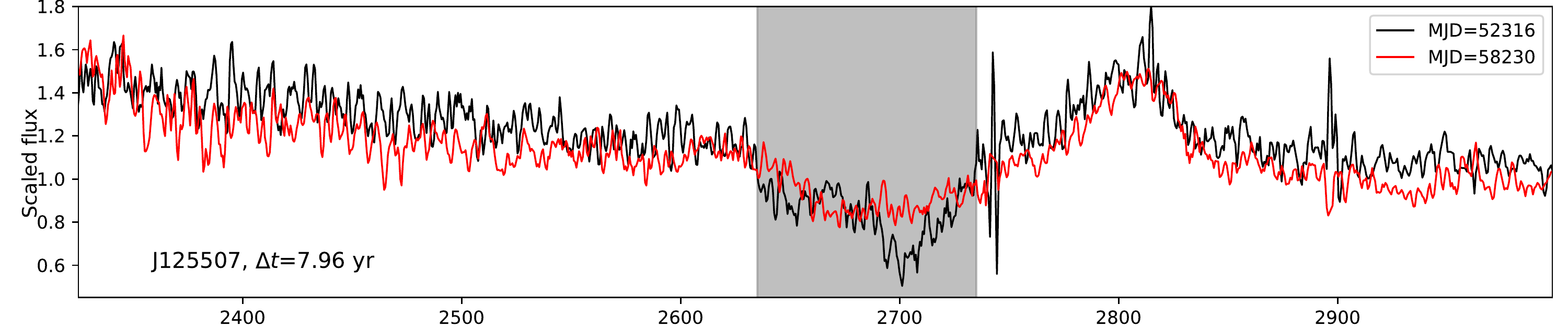}}
\resizebox{6in}{!}{\includegraphics{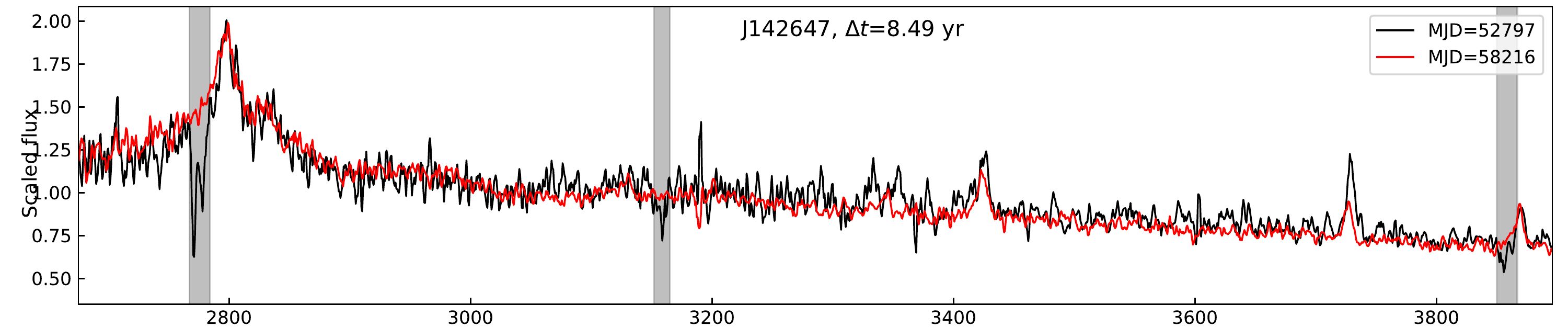}}
\resizebox{6in}{!}{\includegraphics{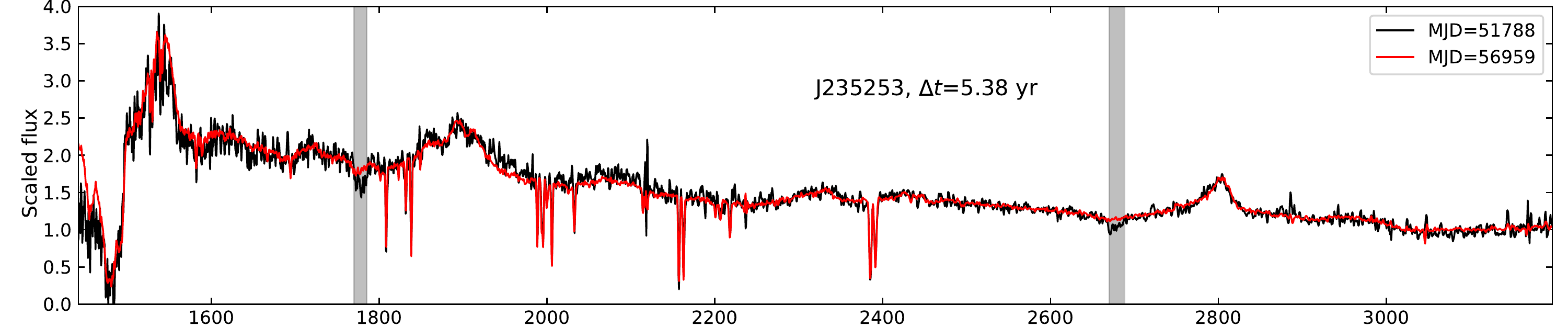}}
\resizebox{6in}{!}{\includegraphics{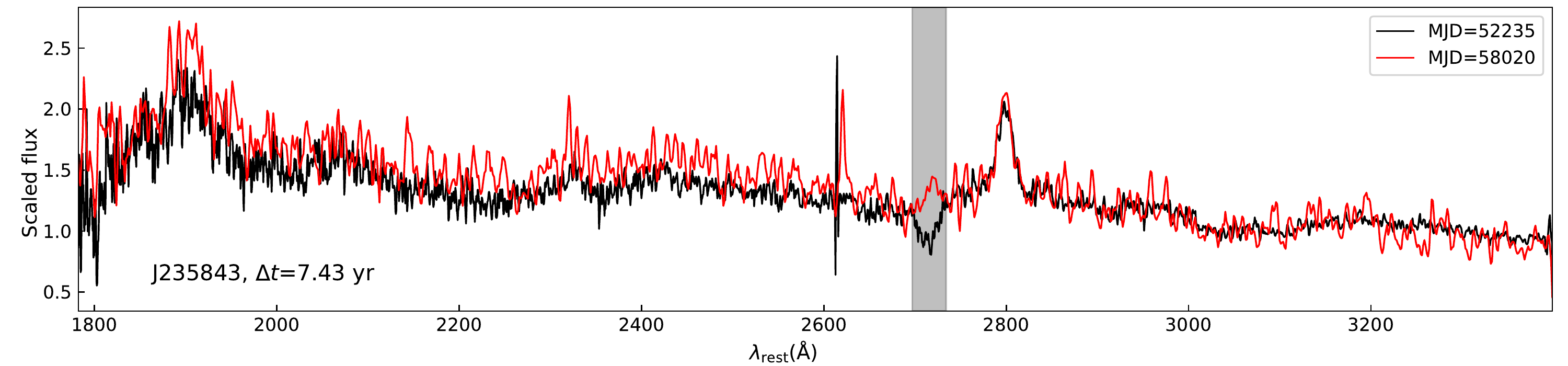}}
\caption{ Demonstration of pristine (J082747, J142647, J235253) and tentative (J1090825, J125507, J235843) BAL transitions (gray/red shadings for BAL disappearance/emergence regions) between two epoch spectra (black and red lines; smoothed for clarity) scaled to the 3000 \AA\ flux density for each quasar.  The cross symbol in the top panel marks a telluric feature. Dashed vertical lines depict the expected wavelengths for \siiv, \civ,  and \mgii\ emission line peaks. $\Delta t$ is the rest-frame sampling interval between the two epochs.  }
\label{f:fig1}
\end{figure*}

The HET/LRS-2 data were processed with the LRS2 pipeline (\citealp{Indahl2019}), which includes standard long-slit spectral extraction procedures, but does not yet include flux calibration. The lack of flux calibration for the HET/LRS2 spectra does not affect the identification of BAL disappearance or emergence. To make the HET/LRS2 spectrum consistent with the nearest-epoch SDSS spectrum for the same quasar, we first fit a low-order polynomial to the spectrum and then scaled the spectrum so that the polynomial fit matches the reddened power-law fit for the SDSS spectrum. In this process, we assumed that the HET/LRS2 spectrum and the corresponding SDSS spectrum have the same continuum  shape across an equivalent wavelength range. 
The YFOSC data were reduced using standard IRAF routines, including bias subtraction, flat field correction, cosmic ray removal, spectral extraction, wavelength identification, and flux calibration. 
The Gemini/GMOS data were reduced using the IRAF package provided from the Gemini Observatory, supplemented by our own tools.

Following \citet{Yi19a}, the minimum and maximum velocities ($v_{\rm min}$ and $v_{\rm max}$) across the entire BAL trough were measured by the outflow velocity where the normalized flux reaches 90\% of the continuum flux on either side of the absorption trough; $v_{\rm cen}$ is the flux weighted velocity center of the absorption trough. Some physical properties, such as luminosity, BH mass, and Eddington ratio, are obtained directly from the SDSS quasar catalogs (\citealp{Schneider10, Shen:2011}).

\section{Spectral measurements and BAL-disappearance/emergence criteria} 
The spectral fitting and related analysis processes of the LoBAL quasars have been described in detail in \citet{Yi19a}. Here, we briefly introduce the main steps. First, a reddened power-law fit was performed to the relatively line-free windows for each quasar. Then, the subsequent fitted power-law function was adopted as a benchmark to guide a non-BAL template fit (see \citealp{Yi19a}). For clarity, all the spectra were normalized at 3000 \AA\ with a suitable smoothing. We note that the quasar (J082747) with both BAL disappearance and emergence was reported in \citet{Yi19b}. In this work, we directly adopt the spectral measurements for the six quasars showing BAL transitions from the two previous studies and complement them with additional  spectroscopic observations (the last sampling epoch spectrum for each quasar).

In \citet{Yi19a}, the following three criteria were adopted to search for ``pristine'' variable troughs from a spectroscopic pair: 
(i) have $\sigma_{\rm\Delta EW}>5$, (ii) have at least one variable region in the trough, and (iii) have a $\chi_{1-2}^2$ value larger than 2.3, where $\sigma_{\rm\Delta EW}$,variable region, and $\chi_{1-2}^2$ were the three metrics defined in Section 4 in \citet{Yi19a}. In addition to the three metrics designed for a robust detection of BAL variability, we have developed three criteria to search for BAL transition events from the parent sample:

(1) A transitioning BAL trough in a quasar is required to be free from any BALs or mini-BALs at any sampling epochs except for the first epoch. However, we do allow the BAL trough region in a spectroscopic pair to contain NALs. 

(2) The $p$-value from a two-sample Kolmogorov-Smirnov (K-S) test across the BAL-disappearance trough in a spectroscopic pair is less than $10^{-4}$. 

(3) A pristine BAL transformation is further identified with the aid of at least one subsequent spectrum after the first detection of BAL disappearance or emergence. 

Although there are cases where NALs may also arise from quasar circumnuclear regions, we do not consider them in this work due to the relatively low spectral resolution and the lack of high S/N spectra. Note that whether or not considering NALs would not change the equivalent widths (EWs) significantly for BALs. 
Variations with $\chi^2_{1-2}$ less than 1.5 may be real in some cases but are not robustly determined, given both the statistical and systematic uncertainties, as well as the BAL-complex troughs sometimes varying from epoch to epoch (e.g., \citealp{Filizak12, Yi19a}). The requirements of at least one variable region and $p$-value less than $10^{-4}$ from the two-sample K-S test allows us to detect BAL-disappearance/emergence events from the sample in a robust way. To make sure the above criteria can detect pristine BAL-disappearance cases, we finally verified all the candidates through visual inspection.

 \begin{table*}
 \center
\caption{Log of observations for the six quasars showing BAL disappearance and/or emergence}
\flushbottom
\begin{tabular}{|c|c|c|c|c|c|c|}
\hline
Name &Instrument& Date&MJD&Exposure Time&$\lambda$ Coverage&Resolution \\
&&(DD-MM-YYYY)&&(s)&(\AA)&($\lambda/\Delta \lambda$) \\
\hline	
                    & SDSS        & 23-12-2001 & 52266 & 9120 & 3800-9200 & 1800  \\
		    & SDSS        & 28-02-2008 & 54524 & 3503 & 3800-9200 & 1800  \\
		    & SDSS        & 11-13-2010 & 55513 & 4504 & 3600-10300 & 1800 \\	
J082747.14+425241.1 & SDSS & 02-10-2015 & 57063 & 5400 & 3600-10300 & 1800 \\
	            & HET/LRS-2 & 04-04-2018 & 58212 & 2700(B) &	3700-6840 & 1800 \\
		    & Gemini/GMOS  & 01-03-2019 & 58486 & 5400(B)/3600(R) & 3700-10350 & 1200 \\

\hline
		    & SDSS        & 01-15-2001 & 51924 & 4500 & 3800-9200 & 1800 \\ 
J090825.06+014227.7 & SDSS  & 12-10-2010 & 55540 & 4504 & 3600-10300 & 1800  \\
		    & HET/LRS-2 & 03-19-2018 & 58196 & 1500(B) & 3700-6840 & 350  \\
\hline 
		    & SDSS        & 02-11-2002 & 52316 & 2702 & 3800-9200 & 1800 \\ 
J125507.12+634423.8 & SDSS  & 02-15-2002 & 52320 & 2554 & 3800-6840 & 1800  \\
		    & SDSS        & 06-01-2013 & 56444 & 3603 & 3600-10300 & 1800 \\
		    & HET/LRS-2 & 04-22-2018 & 58230 & 1500(B) & 3700-6840 & 1800 \\
\hline 
		    & SDSS        & 06-07-2003 & 52797 & 2220 & 3800-9200 & 1800 \\ 
 J142647.47+401250.8 & SDSS  & 04-21-2012 & 56038 & 3603 & 3800-6840 & 1800  \\
		    & SDSS        & 05-16-2012 & 56063 & 2702 & 3600-10300 & 1800 \\
		    & HET/LRS-2 & 04-08-2018 & 58216 & 1200(B) & 3700-6840 & 1800 \\
\hline
                    & SDSS        & 09-01-2000 & 51788 & 1800 & 3800-9200 & 1800  \\
		 & SDSS      & 09-06-2002 & 52523 & 2803 & 3800-9200 & 1800  \\
J235253.51$-$002850.4 & SDSS        & 10-02-2010 & 55471 & 7206 & 3600-10300 & 1800 \\
		    & SDSS        & 10-29-2014 & 56959 & 12601 & 3600-10300 & 1800  \\
		    & HET/LRS-2 & 08-10-2019 & 58705 & 1200(B) &	3700-6840 & 1800 \\
		    & HET/LRS-2 & 08-27-2019 & 58722 & 1200(B) &	3700-6840 & 1800 \\
\hline
		    & SDSS        & 11-22-2001 & 52235 & 5403 & 3800-9200 & 1800 \\ 
J235843.48+134200.2 & SDSS  & 12-04-2012 & 56265 & 2702 & 3600-10300 & 1800  \\
		    & LJT/YFOSC & 09-24-2017 & 58020 & 1800 & 3700-8700 & 350  \\
\hline
\end{tabular}
 \label{table1}
\end{table*}

\begin{table*}
\centering
\caption{Properties of the six quasars undergoing pristine (diamond) and tentative (square) BAL transformations }
\label{tab:pro}
\begin{tabular}{lccccccccccc}
\hline
BAL quasar & $z$ & $M_i$ & log $L_{\rm{bol}}$ & log $M_{\rm BH}$ & log$\lambda_{\rm Edd}$  & $f_{1.4\rm{GHz}}$ & $v_{\rm{centroid}}$  & Ion species of BAL \\
&& (mag) & (erg s$^{-1}$) & ($M_{\odot}$) &  & (mJy) & (km s$^{-1}$) &   \\
\hline
J082747 $\lozenge$  & 2.038 & --25.834 & 45.919 & 9.8 & --1.98 & --1 & --20000; --28300  & Mg II/ Al III/ C IV/ Si IV \\
J090825 $\square$  & 1.002  & --25.347 & 46.28 & 8.76 & --0.58 & --1 & --14000 & Mg II  \\
J125507 $\square$  & 1.034  & --24.799 & 46.07 & 9.15 & --1.17 & --1 & --11000 & Mg II  \\
J142647 $\lozenge$  & 0.749 & --24.183 & 45.857 & 9.39 & --1.64 & --1 & --2773  &  Mg II / HeI*3189/3889 \\ 
J235253 $\lozenge$ & 1.634  & --26.816 & 46.83 & 9.35 & --0.62 & --1 & --13429  &  Mg II / Al III / C IV \\
J235843 $\square$  & 1.134  & --25.671 & 46.39 & 8.78 & --0.49 & --1 & --10100 & Mg II  \\
\hline
\end{tabular}
\end{table*}

\section{Diversity of BAL-profile variability}

\begin{figure*}
\center
\resizebox{1.7in}{!}{\includegraphics{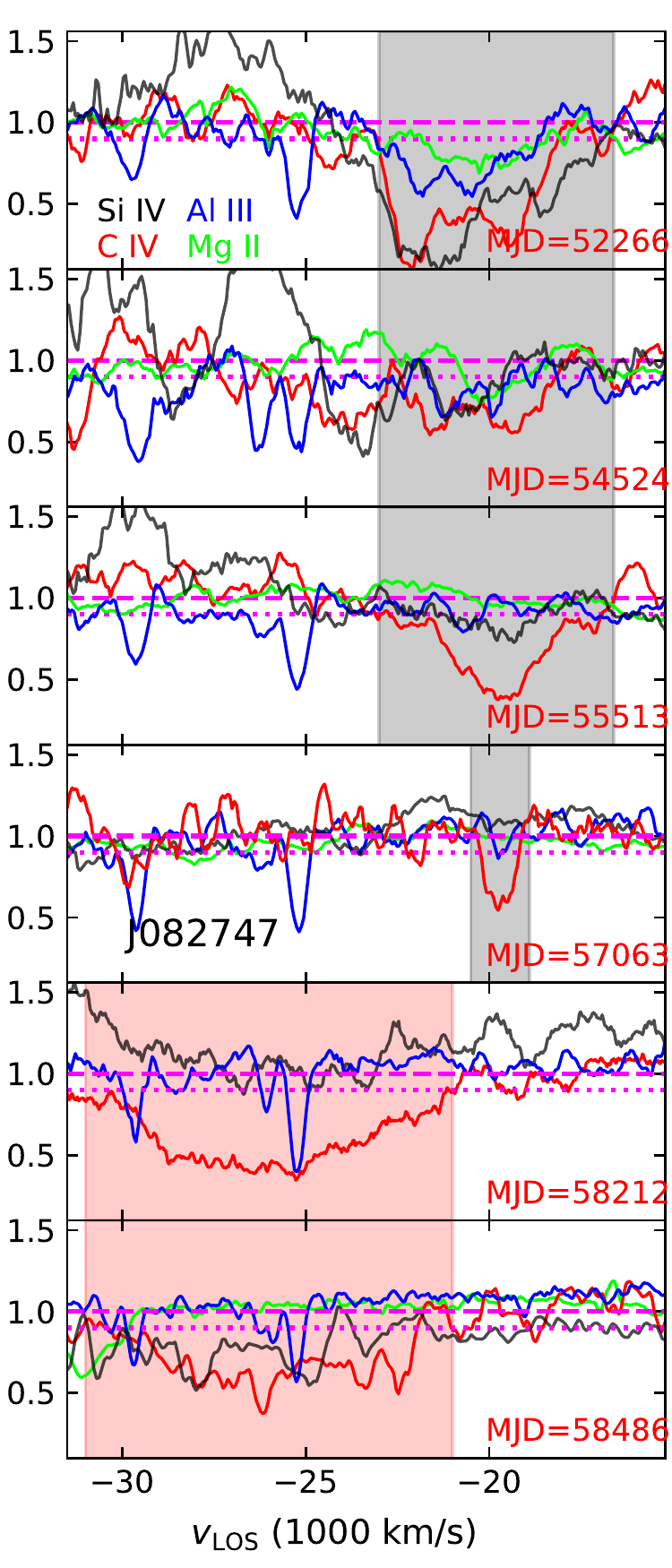}} 
\resizebox{1.7in}{!}{\includegraphics{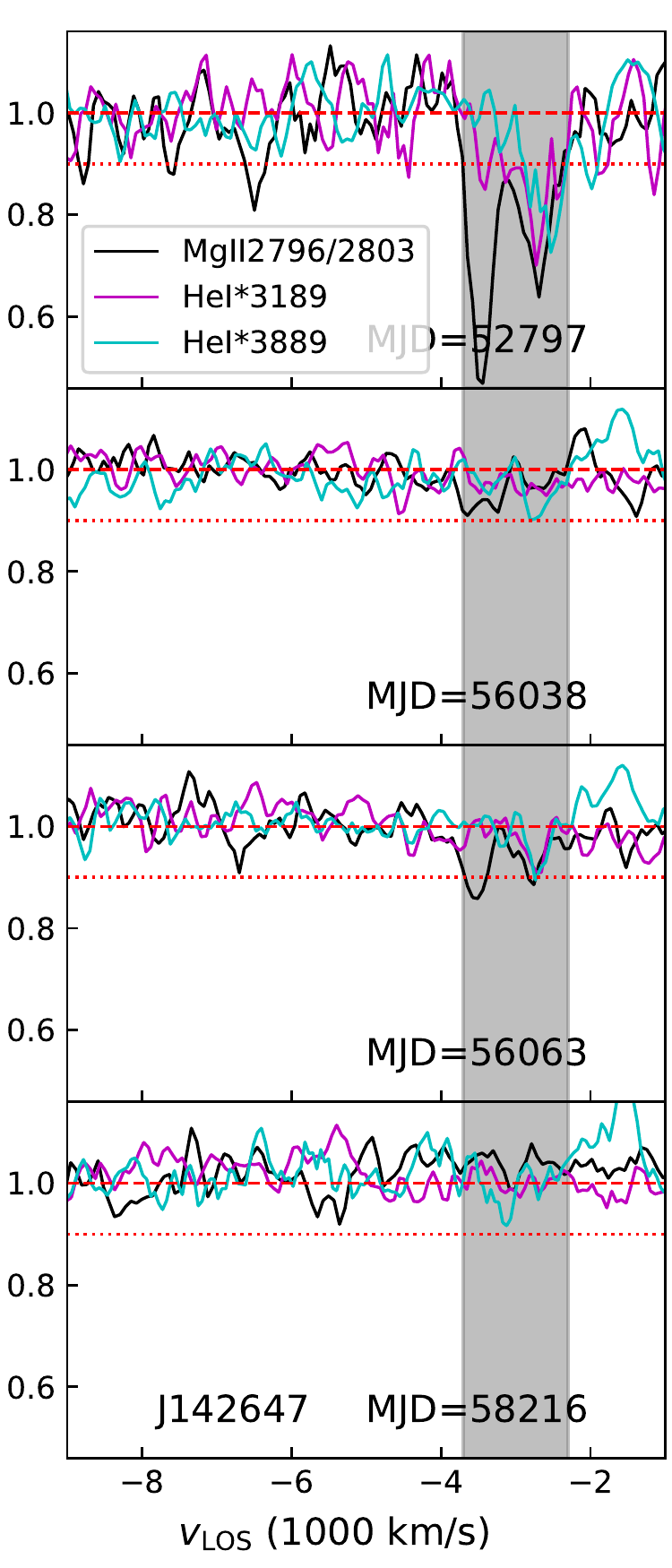}}
\resizebox{1.7in}{!}{\includegraphics{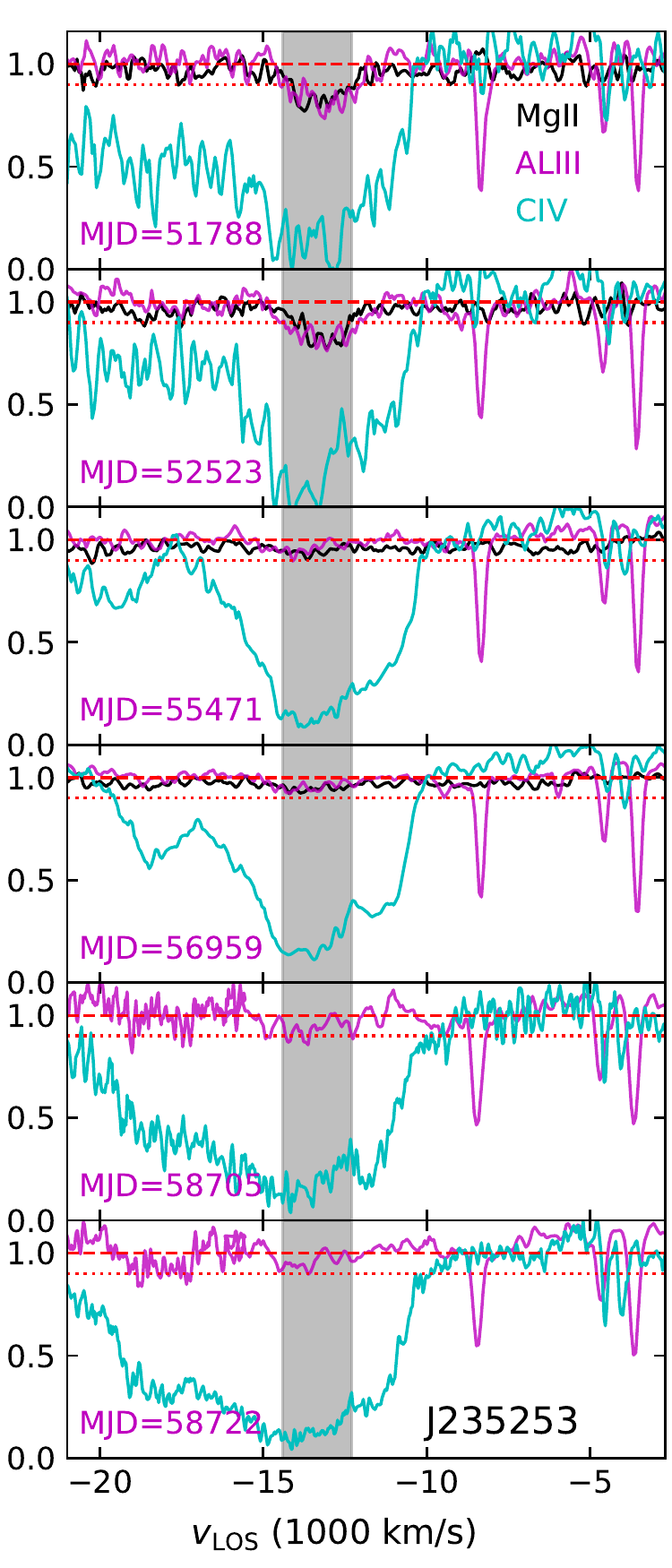}}
\caption{ Top: BAL profile variations (with suitable smoothing for clarity) of the three quasars undergoing pristine BAL transitions. Horizontal  dashed and dotted lines refer to 1.0 and 0.9 normalization levels, respectively.  Gray/red shadings depict BAL disappearance/emergence regions, respectively. All the three quasars show  pristine LoBAL$\rightarrow$HiBAL transformations over the sampling epochs. }
\label{f:fig5}
\end{figure*}

There are six quasars in agreement with all the criteria for a BAL transition, among which we identified three quasars (J082747, J235253, J142647; see Fig.~\ref{f:fig5}) to the pristine BAL transition group. Fortunately, they all have HiBALs and LoBALs at the same velocity, which provides valuable diagnostics to explore the nature of BAL outflows (e.g., \citealp{Arav01,Moe09, Leighly19,Yi19a}). Furthermore, the inner structure and physics of BAL outflows, as well as the main driver of BAL variability, can be well constrained  with the aid of multiple spectroscopic epochs (e.g., \citealp{Vivek18,Yi19b}).

The other three quasars (J090825, J125507, J235843), which are classified to the tentative BAL transition group, have insufficient sampling epochs, relatively low S/N, or possibly contaminated by \feii\ emission variability. Note that their optical spectra cover only \mgii\ and/or \aliii. However, it is worth noting that HiBAL species, typically characterized by deep and wide absorption troughs, most likely exist and overlap the same velocity range to that of LoBAL species based on previous sample studies (e.g., \citealp{Filizak13, Hamann19a, Yi19a}). Despite the absence of HiBAL species from the optical spectra, we can still place some meaningful constraints on the BAL outflows by tracing BAL-profile variability via multi-epoch spectroscopy.

Compared to the sample study of general BAL variability that is  quantified by the EW in \citet{Yi19a}, we focus on individual quasars through the analyses of BAL-profile variability for the six quasars caught in BAL transitions with new observations in addition to the archived SDSS data (see Table~\ref{table1}). To demonstrate the diversity of BAL-profile variability, we present detailed analyses of  the three pristine BAL-transformation quasars with at least four-epoch spectra throughout this Section. The other three quasars undergoing tentative BAL transformations are briefly  described.

\subsection{J082747 and J235253} \label{J082747andJ235253}

J082747 was reported in \citet{Yi19b}, in which its spectral properties and BAL-variability behaviors were analyzed using five spectroscopic epochs (from MJD=52266 to 58212). This quasar shows several prominent features, among which the most dramatic one is the observed BAL disappearance/emergence in multiple ions over the five epochs. Specifically, the low-velocity \civ\ BAL gradually vanished from MJD=52266 to 58212, with large changes both in trough width and depth; meanwhile, a high-velocity \civ\ BAL emerged rapidly between MJD=57063 and MJD=58212. 

The mixed transverse-motion/ionization-change model proposed from \citet{Yi19a} predicts that future observations may detect new BAL emergence in \siiv, \aliii\ and/or \mgii\ at a similar velocity to the high-velocity \civ\ BAL, provided that the newly emerged BAL absorber has the same inner structure and physics to the disappeared BAL absorber. Indeed, the recent spectrum obtained by Gemini/GMOS at MJD=58486  revealed a new \siiv\ BAL emerging  within a very short timescale ($<$100 d in the quasar rest frame) after the \civ\ BAL emergence at the same velocity (see Fig.~\ref{f:fig5});  furthermore, the lack of BAL emergence in \aliii\ or \mgii\  confirms that the low-/high-velocity BAL absorbers most likely arise from two distinct locations along our LOS, as reported in \citet{Yi19b}. The six-epoch sampling data enable the BAL-profile variability to be robustly traced (see Fig.~\ref{f:fig5}). In specific, the low-velocity \civ\ BAL gradually disappeared from MJD=52266 to 58212 while another high-velocity \civ\ BAL emerged rapidly from MJD=57063 to MJD=58212. The simultaneous BAL disappearance/emergence and the obvious differences in kinematics between the low- and high-velocity BAL troughs, together strongly support the aforementioned conclusion that the two different-velocity BALs arise from two distinct absorbers at largely different distances from the quasar center. Since  the analysis of BAL-complex troughs (see Section~3.2 in \citealp{Yi19a}) is sensitive to the number of sampling epochs, it could be ambiguous to identify BAL-disappearance/emergence events via only two-different epoch spectra as demonstrated by Fig.~\ref{f:fig5}. Multi-epoch spectra with well-separated sampling intervals can help to eliminate this ambiguity by tracing BAL-profile variability for individual quasars.

Unlike J082747 mentioned above, J235253 did not show such dramatic BAL-variability behaviors. The \civ-BAL trough is much wider and deeper than \aliii\ and \mgii\ troughs (see Fig.~\ref{f:fig5}), a spectral feature ubiquitously seen in the LoBAL population (\citealp{Yi19a}). 
From visual inspection, the \aliii\ and \mgii\ BALs disappeared at MJD=55471 while the corresponding portion of the \civ\ BAL barely changed over the four SDSS epochs (see the gray region in Fig.~\ref{f:fig5}). This transformation has been further confirmed via two additional spectra obtained by HET/LRS-2 after MJD=56959. Although the two HET spectra do not cover \mgii, they clearly shows the absence of \aliii\ BALs. Thus, it is likely that \mgii\ BALs would not exist over the two epochs, as the relation of BAL variability between them nearly goes hand in hand (\citealp{Yi19a}). Conversely, the \civ-BAL EW in the corresponding velocity range (shaded) did not change significantly over the six epochs, possibly due to saturation when noticing the deepest portion of the entire trough characterized by a nearly flat-bottom shape. However,  the blue wing ($v_{\rm LOS}<-15000$~\kms) of the \civ-BAL trough appears to vary from epoch to epoch. High-quality data are needed before drawing a conclusion concerning this respect. As we did not detect the emergence of LoBAL species in the three epoch spectra after MJD=55471, this quasar is experiencing a LoBAL$\rightarrow$HiBAL transformation. Note that the transformation may be an episodic phenomenon over the quasar lifetime.

J082747 and J235253 show \civ, \aliii, and \mgii\ BALs at the same velocity. Both of them have a similar variability trend in LoBAL species; however, the (\civ) HiBAL-variability behavior between J082747 and J235253 is different. The difference in BAL variability between J235253 and J082747, at first glance, can be explained by different saturation conditions since the \civ\ BAL appears more saturated in J235253 than in J082747.  As shown in Fig.~\ref{f:fig5}, the low-velocity \civ\ BAL (gray region) in J082747 slowly weakened from MJD=52266 to 57063, and eventually disappeared after MJD=58212; while the deepest portion of the \civ-BAL trough (gray region) in J235253 did not change significantly over the six spectroscopic epochs, despite apparent profile changes in the blue wing of the \civ-BAL trough, supporting that different-velocity components in a BAL trough may arise from distinct absorbers at different locations. Further discussion regarding their differences and physical properties is presented in Section~\ref{discussion}.

The BAL-variability behaviors of J235253  clearly demonstrate that the disappearance of LoBAL troughs does not necessarily accompany the disappearance or appreciable variability in HiBAL species at the same velocity. Thus, a reliable identification for BAL$\rightarrow$non-BAL transformation requires long-term spectroscopic monitoring, with a wide-wavelength spectrum covering at least the region between the \siiv\ and \civ\ emission lines.

\subsection{J142647}

\begin{figure*}
\center
\resizebox{1.5in}{!}{\includegraphics{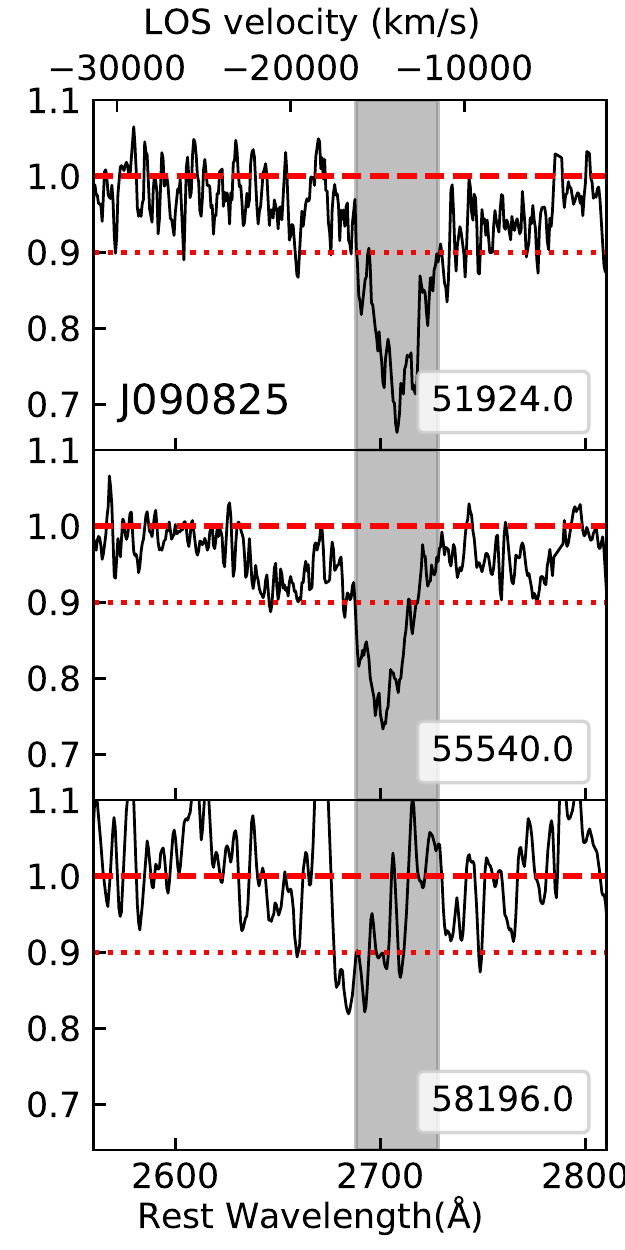}} 
\resizebox{1.5in}{!}{\includegraphics{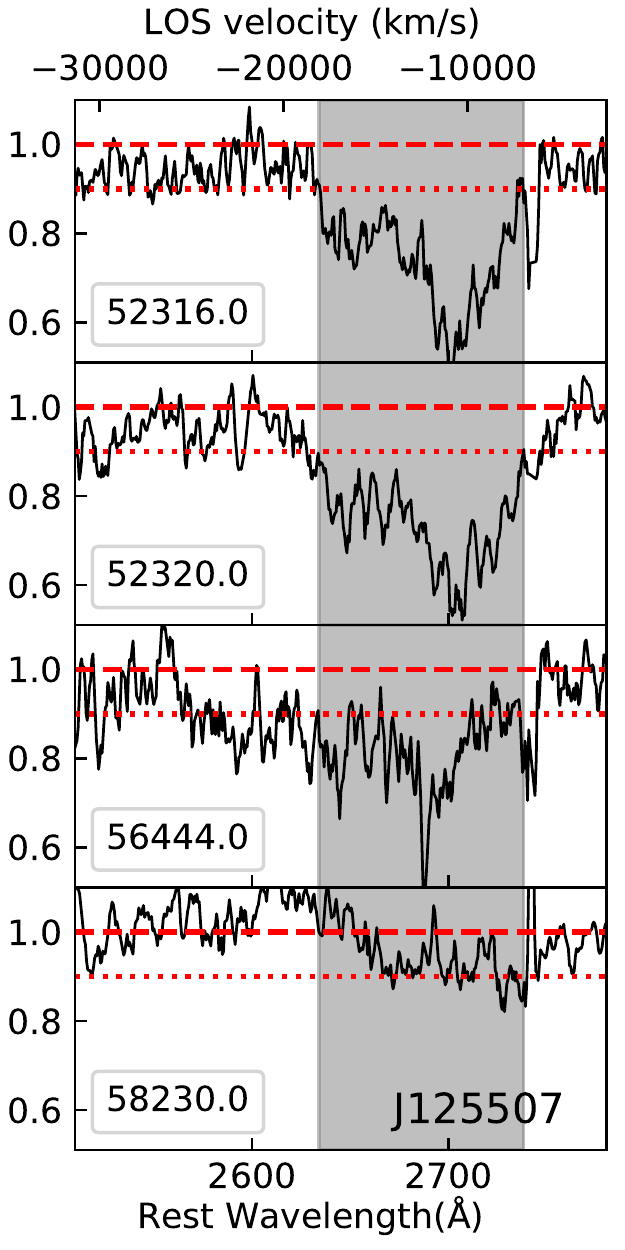}}
\resizebox{1.5in}{!}{\includegraphics{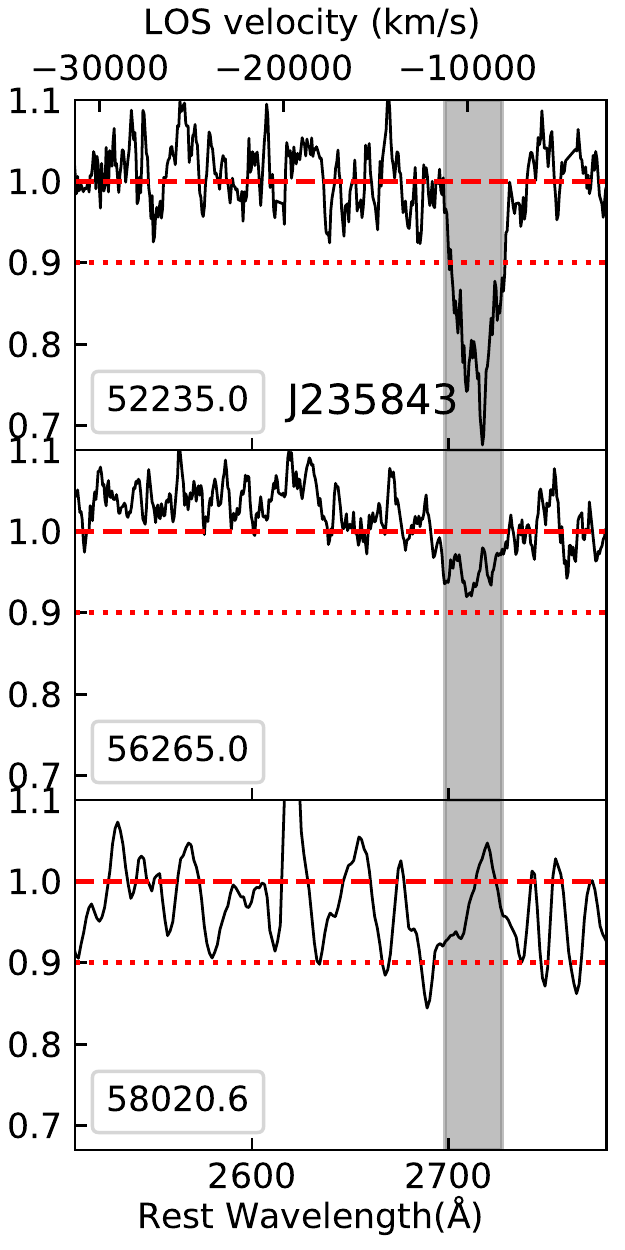}}
\caption{  \mgii-BAL profile variations (with suitable smoothing for clarity) of the three quasars undergoing tentative BAL transitions. Horizontal  dashed and dotted lines refer to 1.0 and 0.9 normalization levels, respectively.  Gray shadings highlight  BAL disappearance regions. We classify them as being in tentative \mgii-BAL transitions because of insufficient sampling epochs, relatively low-S/N spectra in the last epochs, and/or potential contamination from \feii\ emission variability. }
\label{3tentative_transitions}
\end{figure*}
\citet{Liu2015} noted that J142647 shows the most extreme absorption-line variability in their sample, although they did not classify it as a BAL-disappearance event (see Figure A7 in their work), possibly due to the slight residual \mgii\ absorption at MJD=56063. With the aid of a new spectrum obtained by HET/LRS-2, it is clear that the residual \mgii\ absorption completely disappeared at MJD=58216 (see Fig.~\ref{f:fig5}), making it an unambiguous BAL disappearance event. In comparison with the two SDSS spectra at MJD=56038 and MJD=56063, the residual \mgii\ absorption appeared to show a slight change ($\sigma_{\rm \Delta EW}=-3.4$) in trough profile, despite insignificant measured by another metric ($\chi^2_{1-2}=0.9$), tentatively implying a BAL-variability timescale less than 14 days for this quasar. Moreover, the red/blue components of the \mgii\ doublet appear to have a double-peak structure at MJD=56038 and the trough-depth ratio of the \mgii\ doublet is measured to be 1.2 at MJD=56063, again, in support of partial covering toward its background light source by a relatively high optical-depth absorber (\citealp{Hamann19a}).

In addition to \mgii\ absorption, this quasar also shows \hei$\lambda3189,3889$ absorption at the same velocity,  
making it of particular interest, as the \hei$3189,3889,10830$ triplet lines provide ideal diagnostics to explore partial covering and high column-density environments (e.g., \citealp{Leighly11,Liu2015,Sun2017}).  
\hei\ triple lines arise from the same metastable (2 $^3$S) at a rather high excitation level of 19.6 eV,  so \hei\ is essentially a high-ionization line (\hei\ is populated mainly by the recombination process of He$^+$). Noticing that the \mgii\ and \hei$\lambda3189,3889$ absorption lines disappeared from MJD=52797 to MJD=58216, this quasar therefore  supports a similar degree of variability for both \hei\ and \mgii\ absorption lines, perhaps suggesting the lack of hydrogen ionization front in the absorber. This result is consistent with \citet{Leighly19}, where they presented a detailed analysis of the general BAL variability (without BAL disappearance or emergence) in \hei\ and \mgii\ in a LoBAL quasar.

Regarding the BAL profiles of J142647, the spectrum at MJD=52797 did show the doublet of \mgii\ $\lambda\lambda2796,2803$ with a velocity separation of 769$\pm$70 \kms\ (see Fig.~\ref{f:fig5}), probably indicating an unsaturated \mgii\ trough. However, the apparent trough-depth ratio of the \mgii\ doublet is measured to be 1.5, a value between optically thin and optically thick cases. As noted by \citet{Hamann19a}, an observed depth ratio between 1:1 and 2:1 could be difficult in explanation, as it can be caused by either a moderate optical depth or heavy saturation ($\tau>>1$) in an inhomogeneous gas system with velocity-dependent partial  covering factors along our LOS. On the other hand, we found tentative evidence that the red component ($\lambda$2803) of the \mgii\ doublet may be mixed with additional velocity components, when noticing it has a wider wing profile than that of the blue component of the doublet (see the right panel of Fig.~\ref{f:fig5}). Indeed, an approximately equal optical-depth ratio between \hei$\lambda3189$ and \hei$\lambda3889$ in the spectrum at MJD=52797 lends additional support to the partial covering scenario.

The \hei\ triplet lines are rarely seen in the H II region since they come from metastable helium (e.g., \citealp{Leighly11}), therefore the presence of \hei\ may indicate an ionization front existing inside the BAL absorber. Likewise, the \mgii\ ion only starts to become commonplace close to or beyond the hydrogen ionization front (see Fig.~10 from \citealp{Lucy14}). 
Thus, the presence of \hei\ may indicate a high-column density outflow, as reported from previous studies (\citealp{Leighly11, Hamann19a}). However, the similar BAL-profile variability between \mgii\ and \hei\ appears to disfavor the existence of hydrogen ionization front inside the BAL absorber in the context of ideal photoionization simulations (e.g., \citealp{Arav01,Leighly19}). A dedicated analysis including additional data (particularly high-resolution spectroscopy) and sophisticated modeling are required to draw a firm conclusion, which is beyond the scope of this work.

Finally, we note again that, without investigations of \civ\ from UV spectroscopy, it is incapable to establish whether this quasar is transforming from a LoBAL to a non-BAL quasar, although no detection of BAL emergence in \mgii\ and \hei\ at MJD=58216 supports a LoBAL$\rightarrow$HiBAL transformation over the epochs.

\subsection{ Tentative BAL transformations}
There are  three quasars from the sample of 134 quasars showing large amplitude/fractional changes in EW and having little residual troughs in the last epochs. We identified them as being in tentative BAL transformations (see Fig.~\ref{3tentative_transitions}), because (1) two of them have low-S/N spectra in the last epochs; (2) they did not have sufficient sampling epochs to trace the detailed BAL-profile variability; (3) some of them may suffer from \feii\ emission variability, for which we cannot assess from the current data that how much they have been contaminated; (4) unlike the three pristine BAL-transformation quasars, their optical spectra do not cover other ionic BAL troughs at the same velocity that can help to eliminate the ambiguity for BAL transformations. Additional observations are required to monitor their subsequent BAL variability before drawing a conclusion in BAL transformation.

\section{ The incidence of \mgii-BAL transition  and the  \mgii-BAL lifetime }

We clarify that each quasar in \citet{Yi19a} were required to contain \mgii\ BALs in the first epoch spectrum for the construction of the sample. Although the selection effect was considered to slightly bias the frequency of general BAL variability from a large sample (see Section~6.1 in \citealp{Yi19a}), it could significantly or dramatically bias the BAL-emergence rate toward low. This can be easily understood since  (1)  BAL emergence events are rare, making the BAL-emergence rate sensitive to the selection effect; (2) each quasar in \citet{Yi19a} already  contained at least one \mgii\ BAL in the first epoch spectrum and thus the possibility of other distinct BALs emerging in the later epochs would be very low. Therefore, the lack of BAL-emergence detections in the sample could be simply due to the selection effect.  Nevertheless,  we focus on BAL-disappearance events  in the sample. Investigations of the emergence of LoBALs using HiBAL or non-BAL quasar samples are beyond the scope of this work. 

\subsection{The frequency of \mgii-BAL transition }

The \mgii-BAL variability sample in \citet{Yi19a} has delivered the largest such sample ever studied to date, which is the best choice we can use to investigate the frequency of \mgii-BAL transitions. Like the HiBAL variability samples from \citet{Filizak13} and \citet{DeCicco18}, the \mgii-BAL variability sample of \citet{Yi19a} was also constructed from \citet{Gibson09}, making the comparisons between them meaningful. Based on  3/3 quasars exhibiting pristine/tentative \mgii-BAL transitions in the sample of 134 quasars, the corresponding frequency can be derived to be 2.2$_{-1.2}^{+2.2}$\% either in pristine or tentative transformation (the 1$\sigma$ error bounds are calculated following \citealp{Gehrels1986}). 

While the derived frequency of \mgii-BAL transformations in our sample is close to those of \civ-BAL transformations from HiBAL samples (\citealp{Filizak12,DeCicco18}), the frequency of general \mgii-BAL variability that includes all cases with significant variability (see \citealp{Yi19a}) is about half that of HiBAL variability in \citet{Filizak13}. As all the three samples above are constructed from \citet{Gibson09} with similar criteria regarding the detection of BAL variability, we speculate that BAL transitions and general BAL variability may be driven by different mechanisms or BAL transitions are predominantly controlled by the same underlying mechanism, irrespective of specific ionic troughs. We discuss these possibilities in Section~\ref{discussion}.


\subsection{Estimate of \mgii-BAL  lifetime } \label{mgii-BAL-lifetime}
Throughout this work, we define the BAL  lifetime ($t_{\rm BAL}$) as the characteristic time over which a BAL trough is seen along our LOS. For a  large and uniform sample, the average  BAL lifetime ($\langle t_{\rm BAL} \rangle$) of a specific ionic (\mgii) BAL can be  constrained from all BAL-disappearance events in the sample. Note that $\langle t_{\rm BAL} \rangle$ does not necessarily represent a typical real lifetime of these absorbers. Multiple-epoch spectroscopy of a large BAL sample offers an efficient and straightforward way to constrain $\langle t_{\rm BAL} \rangle$ with the aid of BAL transitions (\citealp{Filizak12}). 

\mgii-BAL transitions  are rare events. Only 3 out of 134 quasars are identified to show pristine \mgii-BAL disappearance and no \mgii-BAL emergence events have been found over the sampling epochs from the \mgii-BAL sample of \citet{Yi19a}. 
BAL transition is often thought to be caused by gas absorbers crossing our LOS, changes in the incident ionizing flux, or a combination of the two. 
If the main driver of BAL transition is established for an individual quasar, the observed BAL transition timescale can be further used to constrain the transverse velocity across our LOS and/or the distance of BAL outflow from the quasar center (\citealp{McGraw2017,Yi19b}). 
Regardless of the specific mechanisms in driving BAL variability, $\langle t_{\rm BAL} \rangle$ can be constrained following  \citet{Filizak12}
\begin{equation}
f = <\Delta t>/ <t_{\rm BAL}>
\end{equation}
(see section 4.1 from \citealp{Filizak12}), where $f$ is the fraction of BALs showing transitions over the sampling epochs and $<\Delta t>$ is the average timescale of all observed BAL transitions in the sample, which is 6.89 yr in the quasar rest frame among the six quasars. Note that $<\Delta t>$ is a lower limit on the timescale for a BAL transition event since a BAL certainly exists before the detection in the first epoch. Adopting the fraction $\sim$4.4\% (6/134)   quasars showing BAL disappearance in the sample, $\langle t_{\rm BAL} \rangle$ is then estimated to be $>$160 yr. 
Previous studies reported such a characteristic lifetime of HiBALs  ranging from 100 to 1000 yr from large samples with BAL disappearance/emergence events using the same method (e.g., \citealp{Filizak12, McGraw2017, DeCicco18}). Our estimated characteristic lifetime of \mgii\ BALs is therefore consistent with these studies that are based on HiBAL samples.

Estimating the real lifetime of an entire BAL phase, however, could be extremely challenging due to the unknown structure, geometry, and physics for the BAL wind that is outflowing along many sight lines. 
The current data did not show any \mgii-BAL emergence events in the sample of \citet{Yi19a} and cannot constrain the coasting timescale over which the BAL profile and velocity remain generally unchanged; besides, it is unknown how many BAL episodes would occur over a typical quasar lifetime of order 10$^7$ yr. Therefore, $\langle t_{\rm BAL} \rangle$ estimated above is a conservative lower limit on the real lifetime of \mgii\ BALs. 
 

\subsection{ Variability in continuum shape during BAL transitions } 
Dust variability is of particular interest during BAL transitions, as it can provide valuable diagnostics to investigate BAL properties and hence pin down the origin of BAL variability. Although mid-IR photometry is an ideal tracer of hot dust emission produced by the reprocesses of ionizing photons from the quasar center, many quasars in the LoBAL sample do not have significant detections in W3 and W4 bands from the \textit{WISE} sky survey (\citealp{Wright2010}). Interestingly,  the W1/W2 light curves of the six quasars did not show significant variations. This result is somewhat expected since hot dust emission signal a bulk effect from a circumnuclear region outside the BLR, in which one may not see significant variability over decades. However, LOS dust variability could occur in many cases such as BAL winds breaking out a heavy dust cocoon or entraining interstellar medium (ISM) along specific sight lines. We thus choose the relative variability in UV/optical continuum shape to explore dust variability along our LOS during the BAL transitions. We define  
 \begin{equation}
\Delta \alpha_{\rm 12} =  {\rm log}  \frac{ (f_{5500\rm \AA} / f_{8000\rm \AA} )_2} { (f_{5500\rm \AA} / f_{8000\rm \AA})_1}  
\end{equation}
where $(f_{5500\rm \AA} / f_{8000\rm \AA} )_1$ and $(f_{5500\rm \AA} / f_{8000\rm \AA} )_2$ are the observed-frame  5500~\AA\ and 8000~\AA\ continuum flux density ratios at the early and late epochs, respectively. Consequently, positive/negative $\Delta \alpha_{\rm 12}$ are for being bluer/redder in the later epoch. The two specific wavelengths are chosen to avoid potential flux-calibration uncertainties at the blue or red end of a BOSS spectrum that was not corrected from the pipeline (\citealp{Paris17}), as we did see such cases in a few quasars from the sample. Note that $\Delta \alpha_{\rm 12}$ reflect the FUV/NUV or NUV/optical continuum shape variability in the rest frame, depending on the systemic redshift. 

\begin{figure}
\resizebox{3.4in}{!}{\includegraphics{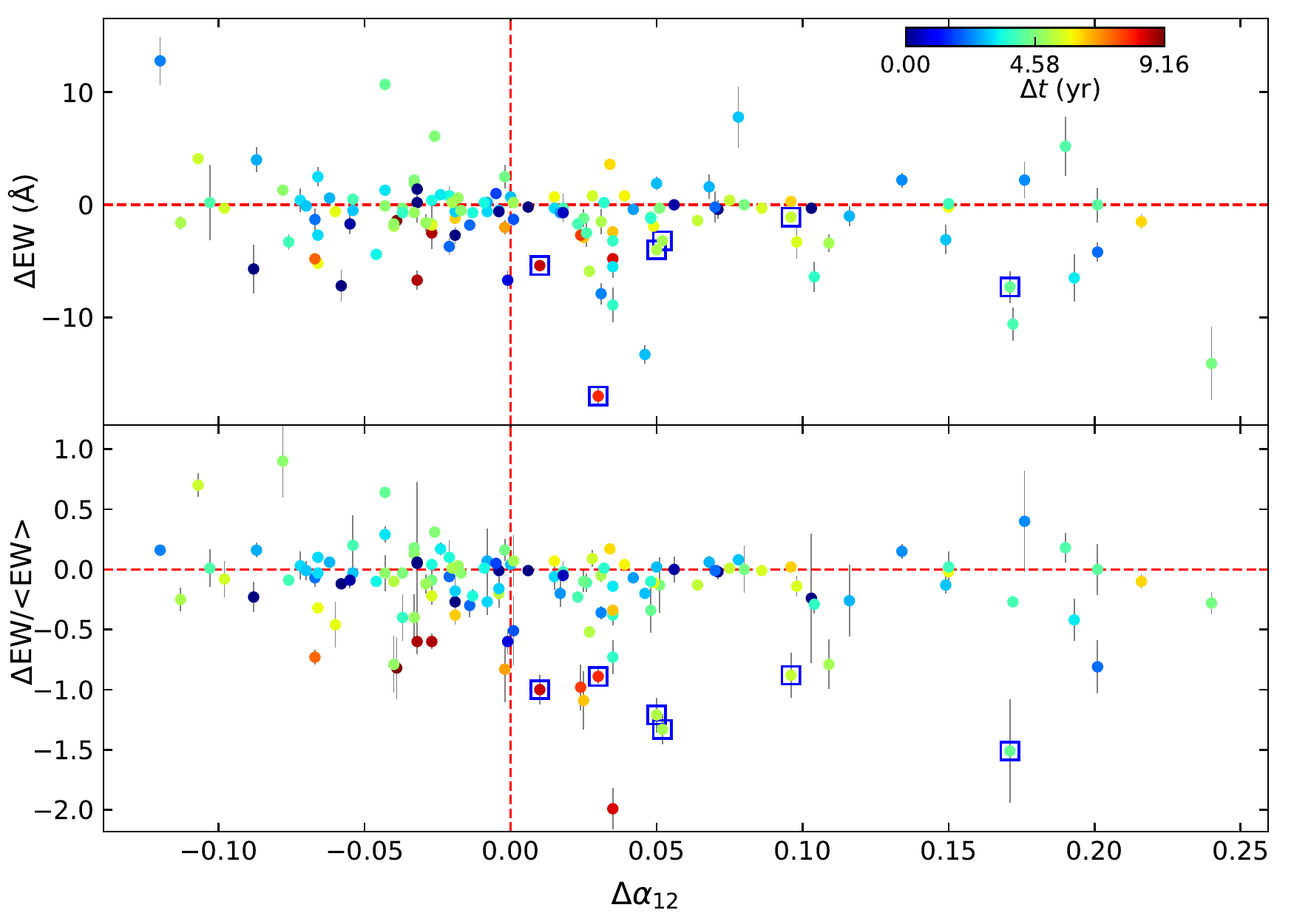}}
\caption{ Distribution of \mgii-BAL variability versus $\Delta \alpha_{\rm 12}$ for the sample consisting of 134 LoBAL quasars. For each quasar, only the spectroscopic pair consisting of the first/last spectra was chosen. The color scheme  displays the rest-frame sampled interval of the spectroscopic pair. The six quasars with BAL transitions are overplotted with squares, among which their continuum shapes become bluer ($\Delta \alpha_{\rm 12}>$0) in  later epochs.  } 
\label{7trans_in134BAL_ERV}
\end{figure}

The distributions of  $\Delta \alpha_{\rm 12}$ versus amplitude/fractional BAL variations for the LoBAL sample are shown in Fig.~\ref{7trans_in134BAL_ERV}. Interestingly, all of the six quasars with BAL transitions (overplotted with squares) appear to become bluer in the later epochs, suggesting a decrease of dust at least along our LOS during the BAL transitions. This result is in agreement with the recent study of quasars showing HiBAL appearance/disappearance (\citealp{Mishra2021}); in addition, \citet{Mishra2021} found a trend that quasars with BAL appearance/disappearance is accompanied by continuum dimming/brightening, implying a ``bluer when brighter'' trend in continuum variability for quasars showing BAL disappearance (see Fig.~8 in their work). This trend can be explained by the dusty outflow models (e.g., \citealp{Zhang2014,Gaskell18}), presumably because changes in dust properties dominate the continuum shape variability, i.e. dust could be destructed when the continuum increases, leading to a bluer continuum shape. Since LoBALs are almost certainly associated with dust given high reddening ubiquitously seen in the LoBAL sample, the bluer-when-brighter trend likely exists among them, particularly when noticing that their continuum shapes become bluer at later epochs during the LoBAL transitions. Evidently, Fig.~\ref{7trans_in134BAL_ERV}  reveals an asymmetry of EW variability at $\Delta \alpha_{\rm 12}>$0, the threshold for quasars being bluer in the later epochs. Therefore, the remarkable time-dependent asymmetry of LoBAL variability reported in \citet{Yi19a} likely signifies an evolutionary effect and the observed LoBAL transitions may be linked to  dust/gas evacuation occurring at least along our LOS. We discuss this  scenario below.  

\section{Discussion} \label{discussion}

We revisit the model invoked  from \citet{Yi19a} to explain the time-dependent asymmetry of \mgii-BAL variability, and comment on how the field has evolved and what have been improved. In the context that LoBALs tend to have relatively high optical depths and low LOS covering factors,  changes in covering factor presumably associated with transverse motions could dominate the strengthening LoBALs while ionization changes and/or other mechanisms dominate the weakening LoBALs (see Section 6 in \citealt{Yi19a}). Indeed, the BAL transitions of the six quasars can be well explained by the model. The improvement in this work is the in-depth analyses of the connection between ionization changes and covering factor changes; specifically, changes in ionization can lead to changes in covering factor without transverse motions. On the other hand, there are several mechanisms being potentially capable of regulating the number density in an obscurer or in a BAL absorber and hence reproducing the observed phenomena in the six quasars. We discuss these issues below. 

\subsection{The advantage from BAL transitions } 

Ascertaining whether different-velocity components appearing in a BAL trough arise from the same or close region is a key challenge before further investigating the BAL nature. Single-epoch spectroscopy with a sufficient spectral resolution and S/N offers a feasible way to address this challenge. However, obtaining high-resolution and S/N spectra cloud be also a great challenge since the vast majority of quasars do not have bright apparent magnitudes. Although the analysis of BAL-complex troughs (\citealp{Filizak13,Yi19a}) based on multi-epoch spectra can alleviate this issue to some extent, ambiguities may still exist without sufficient spectroscopic sampling epochs to trace the detailed BAL-profile variability, as illustrated in Fig.~\ref{f:fig5}.

Conversely, BAL transitions provide rare and unique opportunities in addressing the above challenge. Compared to general BAL variability that is often characterized by changes only in a portion of an entire BAL trough, BAL transitions allow us to establish that all velocity components across a wide BAL trough arise from the same absorber that may consist of many subunits, greatly alleviating the degeneracies during the analysis. From a physical point of view, BAL transitions usually require a larger amplitude change in the main driver than general BAL variability for an individual quasar, irrespective of the underlying mechanisms. This would benefit the analysis of BAL variability and shed light on the origin of BAL variability in an individual quasar. Technically, an investigation of BAL transition does not need high-quality data and high-cadence sampling, as a complete BAL disappearance or emergence event can be easily detected via only two-epoch spectra with a relatively low spectral resolution and S/N.

Although BAL transitions naturally address the above challenge by establishing that all velocity components across a wide BAL trough arise from the same or a close BAL absorber, its location cannot be constrained without knowing the main driver of BAL transitions in a quasar. This is due primarily to the fact that BALs are notorious for partial covering along our LOS, which is dependent on trough velocity, optical depth, density, ionization state, and wavelength etc (\citealp{Arav01, Leighly19, Hamann19}). As an attempt to break down these degeneracies, we start the analysis by assessing the connection between ionization change and covering factor change with the aid of multi-epoch spectroscopy of BAL transitions. 

\subsection{The connection between changes in ionization parameter and covering factor } \label{connection_U_CF}

BAL variability is most widely attributed to ionization changes caused by variation in the incident ionizing flux or transverse motions of BAL absorbers across our LOS. However, discussions of these two possibilities in the literature are often oversimplified since changes in the incident ionizing flux can lead to changes in LOS covering factor of a BAL absorber without transverse motions. For example, \citet{Hamann12} demonstrated that changes in the incident ionizing flux could decrease the (\civ) ionic column density in BAL absorbers, resulting in a smaller effective covering factor along our LOS without transverse motions. In reality,  a full analysis of changes in covering factor is a complex process, as the background emitting source is most likely partially covered by inhomogeneous obscurers and BAL absorbers, posing great challenges in the measurement of the real ionic column density and structure.

\begin{figure}
\resizebox{3.3in}{!}{\includegraphics{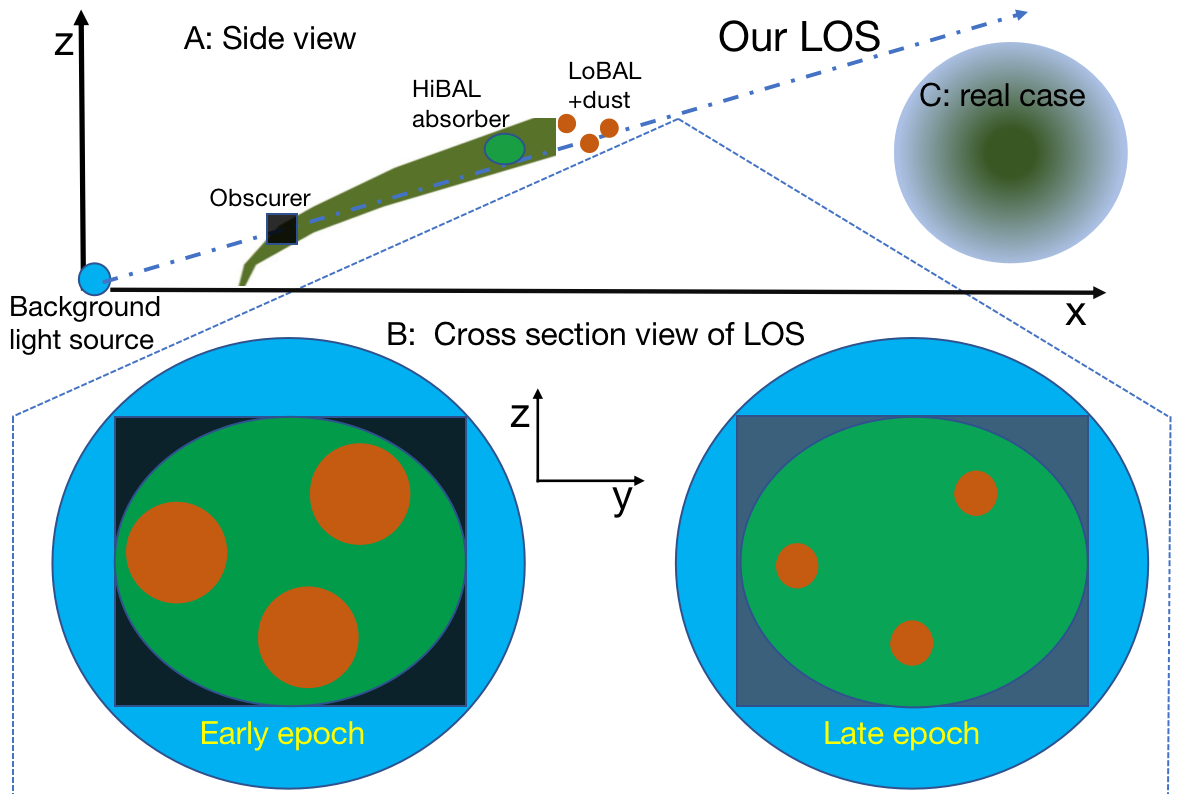}}
\caption{ A schematic illustration of a weakening BAL caused by a decrease in the effective LOS covering factor without transverse motions for the BAL absorber and changes for quasar luminosity. In panel A: the cyan circle, blue square, green circle, and red circles represent the background emitting source, obscurer (a part of the wind base), HiBAL absorber, and LoBAL absorber accompanying dust, respectively (not to scale but father out in order). Panel B demonstrates the cross-section view for these elements along our LOS using effective/homogeneous shapes. The dark to gray colors for the square depict a change of incident ionizing flux between the two epochs, leading to a decrease of   covering factor particularly in the LoBAL absorber.  Panel C shows one of the possible real cases where the obscurer, the BAL absorber, and the dust cloud are all inhomogeneous from a cross-section view.   } 
\label{fig3}
\end{figure}

All the six \mgii-BAL disappearance events can be solely explained by transverse motions of the BAL absorbers accompanying dust. This is particularly true for J082747, as it becomes bluer in the later epochs and possesses two distinct BAL components showing different BAL variability behaviors. Additionally, if \civ\ and \mgii\ BALs co-exist at the same velocity in the same quasar as demonstrated in J082747 and J235253, it is likely that the \mgii\ absorber  associated with dust (see the red circles in Fig.~\ref{fig3}) is embedded inside the \civ\ absorber (green circle) and has a smaller effective covering factor. Although transverse motions alone can explain BAL transitions of the six quasars to a large extent, the current data  cannot rule out the possibility of ionization change as the main driver of BAL transitions for the other five quasars. Importantly, a combination of different mechanisms can  explain the  remarkable time-dependent \mgii-BAL variability trend discovered from the sample, such that weakening BALs outnumber strengthening BALs more evidently on longer sampling timescales (see \citealp{Yi19a}). Together with the results shown in Fig.~\ref{7trans_in134BAL_ERV}, LoBAL quasars indeed favor an evolutionary rather than orientational scenario. 

When revisiting the mixed model proposed in \citet{Yi19a}, an improvement in this work is that transverse motion is not the only mechanism to cause changes in LOS covering factor; in fact, a number of other mechanisms can induce or ``mimic'' changes in LOS covering factor. Throughout this section, we focus on the analyses of the scenarios where changes of the LOS covering factor are caused by changes in the incident ionizing flux density or hydrogen number density without transverse motions, with the aid of BAL transitions detected in the six quasars. We start the assessment with the ionization parameter $U$, which is defined as 
\begin{equation} \label{eq1}
U = \frac{Q_{\rm H}} {4\pi R^2 c n_{\rm H}} \equiv \frac{n_\gamma} {n_{\rm H}} 
\end{equation}  
where $n_\gamma$ and $n_{\rm H}$ are the incident ionizing flux density and hydrogen number density, respectively. It is obvious that changes either in $n_\gamma$ or $n_{\rm H}$ can lead to changes in $U$ and hence determine the fraction of different-ionization species in the absorber (e.g., \citealp{Hamann01, Arav01}). 
As mentioned above,  the observed BAL transitions in the six quasars indicate the presence of a single BAL absorber that  may consist of a number of different-velocity subunits at a similar distance from the quasar center. This absorber represents an extended and perhaps accelerated part of a disk wind as shown in Fig.~\ref{fig3}. The density-change scenario in the BAL absorber itself might be a key ingredient in controlling BAL variability. This is because the quasar center produces enormous radiation, one would expect that ionization parameter $U$ in an outflowing absorber can vary due solely to density changes in the context of radiation pressure confinement, various instabilities, or other compression/dispersion mechanisms during the interaction between BAL wind and ambient ISM (\citealp{Faucher2012a,Wagner13,Baskin14}).

\subsubsection{Changes in the incident ionizing SED } 
Throughout this subsection, we discuss the changes in $n_\gamma$ and keep $n_{\rm H}$ constant within the context of the wind-based unification model proposed by \citet{Dehghanian19b}. The obscurer, as defined in the end of Section~\ref{introduction}, arises from the equatorial base of a disk wind and plays a filtering or shielding role for the transmitted spectral energy distribution (SED) before striking the BAL absorbers (see Fig.~\ref{fig3}). As pointed out by \citet{Dehghanian19b}, the transmitted SED is sensitive to density changes of the obscurer since the Helium or Hydrogen ionization front (see Fig.~3 in that work) may reach or leave the BAL absorber and hence truncate its ionization state. Therefore, a slight change in the obscurer density along our LOS can cause dramatic changes in $n_\gamma$ that would ultimately affect $U$ and kill the specific ion (\mgii) in the BAL absorber, leading to a decrease in the effective (\mgii) covering factor without transverse motions. This scenario was discussed in detail in \citet{Dehghanian19b}. Here, we present an analogical illustration of density changes in the obscurer (dark to gray color changes for the squares in Panel B of Fig.~\ref{fig3}), which leads to changes of the transmitted SED between early and late epochs. As a result, the effective covering factor of the LoBAL absorber associated with dust decreases dramatically in the late epoch. 

Investigating the relation between BAL and continuum variations may offer additional insights into the nature of BAL outflows. Unfortunately, the archived V-band light curves for the six quasars do not allow us to establish the relation between BAL and continuum variations, due to  the lack of photometric observations in the light curves corresponding to the spectroscopic epochs. While all of the six quasars have $\Delta \alpha_{\rm 12}>$0 indicative of being bluer in the later epochs during the BAL transitions (see Fig.~\ref{7trans_in134BAL_ERV}), \citet{Yi19b} revealed that the BAL transition in J082747 ($\Delta \alpha_{\rm 12}$=0.17) is mostly likely dominated by transverse motion rather than ionization change. 
Indeed, the change of incident SED alone cannot explain the observed BAL variability  for J082747, as it contains at least two distinct BAL components characterized by uncorrelated BAL variability behaviors over the sampling epochs, i.e., the simultaneous BAL disappearance and emergence. In addition, BAL and continuum variations may decorrelate when a slight change of density occurs in the obscurer within the context of the wind-based unification model (for details see \citealp{Dehghanian19b}).

The wind-based unification model, when applied to absorption variability cases, predicts coordinated BAL variability in \mgii\ and \civ. Unfortunately, 4/6 quasars cannot be tested with current data due to the lack of \civ\ in the optical spectra. It is worth noting that the BAL variability behaviors in J082747 and J235253 are inconsistent with this model, due to the simultaneous BAL disappearance/emergence and the lack of significant \civ\ BAL variability, respectively. 

\subsubsection{Changes in the absorber density in the RPC scenario} 

The radiation pressure confinement (RPC; \citealp{Baskin14}) model seems appealing to explain the observed results in the six BAL QSOs for three reasons: (1) the BAL-disappearance timescales, the lack of \mgii-BAL emergence events detected in the six quasars, and the remarkable time-dependent asymmetry in \mgii-BAL variability found from the \mgii-BAL sample (\citealt{Yi19a}), are all consistent with the RPC model; (2) A small change of hydrogen column density in a BAL absorber itself can lead to dramatic changes in ionic column densities in the RPC model (see Fig.7 in \citealp{Baskin14}), provided the  ionization front approaches the BAL absorber; (3) RPC can also explain the faster BAL disappearance in \mgii\ than in \civ\ observed in J082747 and J235253 (see Fig.~\ref{f:fig5}), as well as the ubiquity of a stratified ionization structure that is characterized by the co-existence of LoBALs and HiBALs at the same velocity and a shallower trough depth in lower-ionization species in the same spectrum. 

As discussed in Section 4.5 in \citet{Baskin14}, the unique signature predicted from RPC is that when a quasar experiences a transition from a constant low-luminosity to a constant high-luminosity within a few months (a typical value for quasar variability), it would produce a matching rise (a few months) followed by a slow drop of $\sim3$ yr in ionic column density. The 3 yr decline timescale is a lower limit to reach a new steady state for a BAL absorber with a typical hydrogen column density of 10$^{22}$~cm$^{-2}$ and  number density of 10$^8$~cm$^{-3}$ in the RPC scenario. This is consistent with the average BAL-disappearance timescale (6.89 yr) of \mgii\ BALs observed in the six quasars over the sampling epochs and the rapid BAL emergence ($<$ 1~yr in the quasar rest frame) in J082747. A lower-density absorber would lead to a longer decline timescale of the BAL strength. For example, the decline timescale of $\sim$300~yr adopting a typical hydrogen column density of 10$^{22}$~cm$^{-2}$ and number density of 10$^6$~cm$^{-3}$, is consistent with the lower limit of \mgii-BAL lifetime ($>$160 yr) estimated in Section~\ref{mgii-BAL-lifetime}.

However, the hydrostatic RPC solution is valid for non-accelerating outflows and RPC alone cannot explain the observed results in J082747, whose spectra contain at least two distinct BAL components with different BAL-variability behaviors. 
We also point out that the unique signature of a rapid rise and slow drop in ionic column density predicted by the RPC model (see Section 4.5 in \citealt{Baskin14}) may not appear in the case where the incident ionizing SED varies from epoch to epoch.  On the other hand, one should keep in mind that  RPC can occur in a broad range of regions dominated by quasar radiation (typically from 0.01 pc to 10 kpc in a luminous quasar), leading to no constraints on the location of BAL absorbers. In addition, RPC requires an Eddington ratio ($\lambda_{\rm Edd}$) typically larger than 0.1 to ensure the launch of radiation-driven outflows. Although 3/6 quasars have $\lambda_{\rm Edd}$ less than 0.1 at face value (see Table~\ref{tab:pro}), we do not expect a reliable estimate of Eddington ratio via \mgii, as the \mgii\ line may be largely affected by BAL outflows  and internal extinction could be high due to large amounts of dust ubiquitously seen  in the LoBAL population (e.g., \citealp{Urrutia09,Hamann19a,Yi20}). 

\subsubsection{Changes in the absorber density in the BAL-ISM interaction scenario } 
The BAL-ISM interaction scenario, in which high-velocity BAL winds shred, disperse, and sweep up dense clouds, is another promising possibility in producing the BAL transitions observed in the six quasars, given that 5/6 quasars  have BAL velocities larger than 10000 km~s$^{-1}$. As demonstrated in many studies, such a high-velocity wind can cause gas compression and cloud ablation/dispersion/disruption during the expansion (e.g., \citealp{Faucher2012a, Wagner13, Zubovas14, Waters17}), leading to the patchy/filamentary, dusty, and inhomogeneous absorption  along our LOS (\citealp{JiangP2013,Zubovas14, Veilleux16, Gaskell18}). This can be easily understood since a wide opening-angle quasar wind would flood through the intercloud channels, sweep up the ISM, and ablate/disperse the dense cloudlets. In particular, a BAL wind interacts strongly with the inhomogeneous ISM consisting of dense and perhaps dusty cloudlets embedded in a tenuous medium, which, in turn, produces patchy/filamentary obscuration and absorption effects. As a result, many sight lines are filled with numerous small dusty cloudlets like a spray distributed in front of the continuum emitting region (e.g., \citealp{Hamann12,Leighly19}). This scenario may account for high reddening and partial covering in LoBAL/FeLoBAL quasars.

BAL transition offers a unique opportunity to test the above scenario. As mentioned in Section~\ref{mgii-BAL-lifetime}, the mean \mgii-BAL lifetime along our LOS is estimated to be $>$160 yr for the  sample consisting of 134 \mgii-BAL quasars. This does not contradict  the cloudlet survival time (an order of a few thousand yr) mentioned in \citet{Faucher2012b}, in that the BAL lifetime estimated above is the characteristic time of a \mgii-BAL absorber seen only along our LOS, while the cloudlet survival time is the real physical lifetime including the time for a BAL absorber moving out of our LOS. Furthermore, the timescale of cloudlet formation is much shorter than that of cloud disruption (\citealp{Faucher2012a,Faucher2012b}), which can also explain the lack of observed \mgii-BAL emergence events, the high optical depths/low covering factors, and the remarkable time-dependent asymmetry of BAL variability discovered in the \mgii-BAL sample of \citet{Yi19a}. Given that interactions between high-velocity winds and dense clouds can occur either close to the quasar center or much farther out in the host galaxy, the BAL location cannot be constrained by BAL variability alone. However, the BAL-ISM interaction must leave pronounced signatures such as BAL deceleration; when taken as a whole, one can pin down the BAL wind location (Yi et al. in prep).  We do note that, unlike RPC, transverse motions of cloudlets caused by strong interactions between high-velocity winds and dense ISM clouds are naturally expected. Therefore,  the BAL-ISM interaction may be more likely to cause the decrease of dust and/or transverse motions along our LOS at larger radii than other scenarios. 


\subsection{ An overall trend of BAL transformations}

Throughout this work, we clarify that a quasar identified as being in BAL transformation over the observed epochs is different from a BAL disappearance or emergence event since a quasar could have multiple BAL troughs. While the FeLoBAL$\rightarrow$LoBAL transformations have been found in several quasars (\citealp{Hall11,Rafiee16,Stern2017}), the transformation of being a FeLoBAL quasar, to our knowledge, was not reported in any quasars from the literature, which can be attributed to the rarity or equally a short formation timescale of FeLoBALs. Similarly, while pristine/tentative LoBAL$\rightarrow$HiBAL transformations have been detected in 3/3 quasars in the LoBAL sample consisting of 134 \mgii-BAL quasars from \citet{Yi19a}, no \mgii-BAL emergence events have been found in this sample, despite a potentially large bias due to the selection effect. By contrast, the non-BAL$\rightarrow$HiBAL transformations have been reported in a number of quasars over rest-frame 3 yr from \citet{Rogerson18}, although the rate (0.59$\pm$0.12\%) appears  lower than the HiBAL$\rightarrow$non-BAL rate (2.3$^{+0.5}_{-0.4}$\%) over rest-frame 5 yr from \citet{DeCicco18}. Together with the fact that reddening and hydrogen column density decrease dramatically from FeLoBAL, LoBAL, HiBAL to non-BAL populations (e.g., \citealp{Reichard03,Hamann19a}), our investigations suggest an overall FeLoBAL/LoBAL$\rightarrow$HiBAL/non-BAL transformation sequence, in which FeLoBAL and LoBAL quasars are in a relatively short-lived, young phase accompanying dusty outflows. Note that it is unknown how many episodes of such BAL transformations would occur over the entire quasar lifetime. Interestingly, the remarkable time-dependent trend of BAL variability, such that the weakening BALs outnumber strengthening BALs more evidently on longer sampling timescales (\citealt{Yi19a}), mostly occurs in LoBAL quasars being bluer in the later epochs and was not found in the HiBAL population (e.g., \citealp{Filizak13,DeCicco18}). These differences further support that, unlike FeLoBAL/LoBAL quasars at an earlier phase, HiBAL quasars could be at a later phase in which their heavy dust cocoons have been largely cleaned out by LoBAL winds. In addition, previous studies reported that a few quasars have undergone HiBAL$\rightarrow$non-BAL$\rightarrow$HiBAL or non-BAL$\rightarrow$HiBAL$\rightarrow$non-BAL transitions (\citealp{McGraw2017,Rogerson18}), which can be well explained by largely dust-free, optically thin absorbers that are sensitive to small changes in ionization state at a relatively late evolutionary phase.  

The BAL transformation sequence can be better understood in the context of quasar feedback, where previous quasar winds (disappeared now) broke out a heavy dust cocoon and left substantial small clumps around the quasar center. If the quasar continues to launch high-velocity, wide-angle  winds that keep shredding/dispersing/cleaning these dusty clumps out of the quasar host galaxy, a blue and non-BAL quasar will be ultimately uncovered (\citealp{Sanders88,Urrutia09,Hopkins10,Wagner13}). Indeed, the higher BAL detection rate  in red quasars than in blue quasars, the more energetic winds in LoBALs than in HiBALs, the remarkable time-dependent trend of LoBAL variability, and the predominant LoBAL$\rightarrow$HiBAL transformations together strongly favor an early-phase dust evacuation scenario for LoBAL quasars (e.g., \citealp{Urrutia09,Hamann19a,Yi19a}), presumably LoBALs arise from dusty environments. Conversely, the  transformations between HiBAL and non-BAL quasars may be dominated by changes in ionization state for relatively dust-free, optically thin absorbers at the late/blue phase, as reported in recent studies (e.g., \citealp{Wang15,He17,Vivek18,Mishra2021}). On the other hand, mounting observational studies have found that quasar winds can possess different manifestations traced mostly by BALs, mini-BALs, NALs, and/or broad emission-line blueshifts from spectroscopy. Any form of high-velocity quasar winds with high column densities may have the potential in expelling dust/gas from small to large scales (\citealp{Yi20,Xu20a,Xu20b,Paola20,Choi2020}), although the actual impact of feedback strongly depends on the specific environment and wind-ISM coupling process that may vary from object to object.

\section{Summary}

Built upon the sample study of \mgii-BAL variability in \citet{Yi19a}, we further explore the inner structures and physics of these quasars undergoing \mgii-BAL transitions with the aid of new observations. The major conclusions are summarized as follows:

\begin{enumerate}
\item
We identify 3/3 quasars undergoing pristine/tentative BAL transformations (\mgii-BAL disappearance) but no quasars showing \mgii-BAL emergence from 134 LoBAL quasars in \citet{Yi19a}. The incidence is therefore derived to be 2.2$_{-1.2}^{+2.4}$\%  in pristine BAL transformation. 
\item
An average characteristic  lifetime of \mgii\ BALs  along our line of sight is constrained to be $>$160 rest-frame  yr from the largest \mgii-BAL variability sample studied to date (\citealp{Yi19a}). 
\item
Our analyses indicate that changes in ionization parameter and covering factor are inherently connected and cannot be easily distinguished. 
\item
While BAL transitions in J082747 are most likely due to transverse motions,  additional mechanisms like RPC, SED filtering, or BAL-ISM interactions can explain BAL transitions for the other five quasars. However, any individual model is difficult to unify the diversity of BAL-profile changes in the six quasars, particularly when a quasar possesses more than two distinct BAL troughs with different variability behaviors. 

\item
Our investigations of \mgii-BAL transitions, in combination with previous BAL studies, suggest an overall  FeLoBAL/LoBAL$\rightarrow$ HiBAL/non-BAL transformation sequence accompanying a decrease in reddening, consistent with the evacuation models for the origin of commonly seen blue quasars.
\end{enumerate}

Ascertaining whether different velocity components appearing in the same BAL trough arise from the same location is a key step in the analysis of the BAL nature. BAL transition naturally addresses this issue by establishing that all different-velocity components in an observed BAL trough arise from the same absorber, even without high-quality, high-cadence spectroscopic data. However, the inner structure and physics of BAL outflows are more complex than those from previous studies with oversimplified assumptions. Future investigations such as BAL acceleration/deceleration,  relations between BALs and emission lines, spatial information both in absorption and emission, the combination and comparison with other type objects possessing dusty outflows etc., can offer valuable  diagnostics to explore the nature of LoBAL quasars from a comprehensive, in-depth view. Therefore, follow-up, multi-wavelength data covering longer timescales are required to further probe the BAL physics, time evolution, and their potential for quasar feedback. 

\acknowledgments

We thank the referee for comments and suggestions that helped to clarify the manuscript. 
W. Yi acknowledge support from  NSF grant AST-1516784. W. Yi also thanks support from the Chinese National Science Foundation (NSFC-11703076) and the West Light Foundation of The Chinese Academy of Sciences (Y6XB016001), as well as the financial support from the China Scholarships Council (No. 201604910001) for his postdoctoral study at the Pennsylvania State University.  JT acknowledges support from NASA ADP grant 80NSSC18K0878. 

Funding for SDSS-III has been provided by the Alfred P. Sloan Foundation, the Participating Institutions, the National Science Foundation, and the U.S. Department of Energy Office of Science. 
The Low-Resolution Spectrograph 2 (LRS2) was developed and funded by the
University of Texas at Austin McDonald Observatory and Department of
Astronomy, and by the Pennsylvania State University. We thank the
Leibniz-Institut f\"ur Astrophysik Potsdam and the Institut f\"ur
Astrophysik G\"ottingen for their contributions to the construction
of the integral field units. 

This work uses data obtained from the Gemini Observatory (PI: WY; program ID: GN-2018B-FT-214), which is operated by the Association of Universities for Research in Astronomy, Inc., under a cooperative agreement with the NSF on behalf of the Gemini partnership: the National Science Foundation (United States), the National Research Council (Canada), CON- ICYT (Chile), Ministerio da Ciencia, Tecnologia e Inovaciao (Brazil) and Ministerio de Ciencia, Tecnologia e Innovacion Productiva (Argentina).
The Hobby-Eberly Telescope (HET) is a joint project of the University of Texas at Austin, the Pennsylvania State University, Ludwig-Maximillians-Universit\"{a}t M\"{u}nchen, and Georg-August-Universit\"{a}t G\"{o}ttingen. The Hobby-Eberly Telescope is named in honour of its principal benefactors, William P. Hobby and Robert E. Eberly. 
We acknowledge the support of the staff of the Lijiang 2.4m telescope (LJT). Funding for the telescope has been provided by CAS and the People's Government of Yunnan Province.

\end{document}